# THE ROLE OF CHAOS IN THE CIRCULARIZATION OF TIDAL CAPTURE BINARIES. II. LONG-TIME EVOLUTION.

Rosemary A. Mardling

Mathematics Department, Monash University,
Clayton, Victoria, Australia, 3168

*email:* r.mardling@maths.monash.edu.au

## ABSTRACT

A self-consistent, adiabatic model for the *long-time* behaviour of tidal capture binaries is presented. It is shown that most capture orbits behave chaotically, with the eccentricity following a quasi-random walk between the values of $\stackrel{<}{\sim} 1$ and some lower limit associated with the periastron separation at capture.

If dissipation is taken into account, the binary goes through a short and violent chaotic phase, followed by a long quiescent phase in which it slowly circularizes from a high eccentricity on a much longer timescale than previously thought. A consequence is that merger is less likely than previously thought, and hence such binaries will be available as a heat source to the cores of globular clusters, particularly while they are in the less tightly bound, highly eccentric phase.

If the model is correct, any highly eccentric binaries observed in globular clusters which contain a main-sequence star will most likely be found to have a period derivative much smaller than that predicted by the standard model.

We also predict that the companion of PSR B1718-19 in NGC 6342 which is a globular cluster binary likely to have been formed by tidal capture (Wijers & Paczyński 1993) will be found to have a mass of around $0.2M_\odot$.

The model may be used to describe the evolution of low-mass X-ray binaries, pulsar binaries, and cataclysmic binaries which abound in globular clusters.

## 1 Introduction

The perceived role of tidal capture binaries in the evolution of globular clusters has varied considerably in the last few years. Initially it was thought that their formation during and after core collapse would provide the main source of energy (via encounters with single stars and other binaries) for halting any further collapse of the core, and indeed its reexpansion (see, for example, Statler, Ostriker, & Cohn 1986). As work proceeded, it was realized that the extremely close encounters needed to form capture binaries would most probably lead to the merger of the two stars in most cases (McMillan, McDermott & Taam 1987, hereafter MMT, and Ray, Kembhavi & Antia 1987). This is because circularization of the orbit is thought to occur on a timescale much smaller than the dissipation timescale of the stars, and hence the stars' response to the dissipation of the enormous amount of oscillation energy present by the time circularization has taken place is to expand, leading to contact or merger of stars that would not have contacted or merged had they followed a normal evolutionary path. If a binary merges, it is removed as a source of energy for the core, and since so many are thought to suffer this fate, the mechanism is no longer thought viable. It is now thought that the source of binaries is those pre-existing in the cluster, that is, those formed primordially. Despite the difficulties observing them, some binaries containing a giant have been observed in the more dispersed regions of some clusters, and the frequency of occurence of such binaries has been inferred (Hut et al. 1992). It now appears that there may be enough primordial binaries to act as an efficient energy source for the core (Goodman & Hut 1989).

Of the capture binaries that do survive, for example some of those formed by encounters between compact stars and low mass main sequence stars, it is thought that encounters with other stars in the cluster would almost certainly eject them from the cluster, or at least from the core (Goodman 1989). This is because capture binaries are thought to circularize very quickly, and hence spend most of their life in very hard circular orbits. A superelastic encounter with a third star would most likely cause the binary to recoil to the extent that the kinetic energy gained by the binary and the third star at the expense of the orbit would be enough to eject them at least from the core. This process can have observational consequences, for if any of the ejected components was a millisecond pulsar, it may be observed away from the core, where such objects are likely to have been formed. It is likely that the binary pulsar 2127+11C in the globular cluster M15 has suffered this fate (Prince, Anderson, & Kulkarni 1991). On the other hand, the present model suggests that some capture binaries spend a substantial time in highly eccentric, less tightly bound orbits, and hence encounters with a single star may be far less violent.

Even if the tidal capture process is not important for halting core collapse and driving core expansion, a knowledge of the evolution of binaries after capture is necessary to understand the formation and evolution of the many low-mass X-ray binaries (LMXBs), binary pulsars and cataclysmic binaries observed in the cores of globular clusters. The question of whether or not so many encounters do lead to merger may be answered with a dynamical model of the process. One problem with the merger scenario is that it predicts higher core luminosities than are observed (Goodman 1989). This lends weight to the present model, to be discussed shortly.

The only models for the evolution of binaries after capture to date have partially (Kochanek 1992) or wholly (MMT, Ray et al. 1987) depended on the model devised by Press & Teukolsky (1977), which calculates the amount of energy deposited after the *first* periastron encounter. This model assumes the orbit remains parabolic during the encounter. MMT and Ray et al. (1987) assumed that the energy deposited during subsequent periastron encounters was independent of the state of the stars, and hence that the formula given by Press & Teukolsky (1977) could be used, at least for reasonably high eccentricities. This results in extremely short circularization times, of the order of 10 years for extremely close encounters.

On the other hand, Kochanek (1992) predicted that the energy exchange at periastron would indeed depend on the oscillatory state of the stars, and hence that the eccentricity of the orbit would perform a quasi-random walk, a prediction borne out by the results of the present calculations. He used the 'affine' model for extremely close encounters, which all lead to disruption. This model assumes that the structure of the star can be represented by a global distortion, such that the surfaces of constant radius in the initial spherical star are distorted into self-similar ellipsoids. It has the advantage of being self-consistent in the same sense as the present model, and of naturally including non-linear terms so that oscillation amplitudes need not be small, but the disadvantage of only representing the $l = 2$ fundamental mode. Moreover, it is computationally expensive, so that only a few orbits following very close encounters can be studied. For wider orbits, Kochanek (1992) assumes the Press & Teukolsky (1977) formalism is valid, albeit while assuming that the eccentricity follows a quasi-random walk.

The present model is based on a linear, adiabatic normal mode analysis developed by Gingold & Monaghan (1980) and outlined in Mardling (1994, hereafter Paper I). The self-consistent nature of the model allows us to follow the evolution of the system indefinitely (for as long as the energy is conserved to within some tolerance).

In this paper, the model is modified by taking advantage of the fact that for capture orbits, the orbit and oscillations are effectively decoupled for a large part of the orbit (Section 3). This allows us to study relatively wide capture orbits for many thousands of orbits using a reasonable amount of computer time (Section 4). Section 5 considers the response of the system to dissipative processes, while Section 6 contains concluding remarks.

## 2  Tidal Capture Cross Sections

Tidal capture in globular clusters is possible for an extremely small range of capture cross sections: it is unlikely that the right conditions prevail elsewhere in the galaxy. In the following, we calculate the range of periastron separations for which *stable* binaries can result for a range of mass ratios appropriate to the cores of globular clusters, *assuming that one object is compact* and assuming a relative velocity at infinity of 10 km s$^{-1}$. Although this calculation is not new, we present it here to highlight the fact that
1. The periastron separation at which the extended object fills its Roche lobe places a lower limit on stable binary formation,
2. This lower limit increases with increasing mass ratio of compact to extended object, and
3. The range of periastron separations which lead to stable binaries increases with increasing mass ratio, $s$, of compact to extended object, so that the capture process favours lower mass companions.

For the capture process to result in a bound orbit, the pair of stars must surrender to the tides at least as much gravitational potential energy as their relative kinetic energy at infinity. Thus the capture cross section is $\sigma = \pi R_0^2(v_\infty)$, with the velocity at infinity, $v_\infty$, given by

$$v_\infty = \sqrt{2\Delta E/\mu}, \qquad (1)$$

where $\mu = M_1 M_2/(M_1 + M_2)$ is the reduced mass with $M_1$ the mass of the extended star, $M_2$ the mass of the compact object, and $\Delta E$ the energy transferred to the tides after the first periastron passage. The capture impact parameter, $R_0$, is given by (Lee & Ostriker 1986)

$$R_0^2(v_\infty) = 2G(M_1 + M_2)R_{\min}/v_\infty^2, \qquad (2)$$

where $R_{\min}$ is the distance of closest approach. Given $R_{\min}$, we may calculate $\Delta E$ (see Paper I) and hence $R_0$. Typically, this is about 50 times $R_{\min}$ for $M_1 + M_2 = 2M_\odot$ and $v_\infty = 10 \mathrm{km\ s}^{-1}$.

Table 1 presents ranges of periastron separations for stable binary formation (in units of $R_*$, the radius of the extended star), where $R_{\mathrm{RL}}$, the distance at which the extended star fills it Roche lobe, is given by (Sahade et al.)

$$R_{\mathrm{RL}} = \begin{cases} 1/(0.38 - 0.2\log_{10} s) & 0.05 < s < 1.25 \\ 2.165(1+s)^{1/3} & s > 1.25, \end{cases} \qquad (3)$$

| $s$ | $R_{\mathrm{RL}}$ | $R_{10}$ | $\delta$ |
|---|---|---|---|
| 0.5 | 2.27 | 2.82 | 0.55 |
| 1 | 2.63 | 3.19 | 0.56 |
| 1.4/0.7 | 3.12 | 3.72 | 0.60 |
| 1.4/0.4 | 3.57 | 4.30 | 0.73 |
| 10/0.7 | 5.37 | 6.53 | 1.16 |

Table 1: Ranges of periastron separations for stable binary formation for various mass ratios, $s$. The lower limit, $R_{\mathrm{RL}}$, is the distance at which the binary will fill its Roche lobe, while the upper limit, $R_{10}$, is the maximum distance for which bound orbits can result when $v_\infty = 10$ km s$^{-1}$.

$R_{10}$ is the limiting periastron separation for which bound orbits result when the relative velocity at infinity is 10 km s$^{-1}$, and $\delta = R_{10} - R_{\mathrm{RL}}$.

We have included an equal mass pair to illustrate the case of, say, a low mass white dwarf and a low mass main sequence star, two pairs for which the compact object is a neutron star and the other a low mass main sequence star ($0.7 M_\odot$ and $0.4 M_\odot$) and a pair for which the compact object is a $10 M_\odot$ black hole and the other a $0.7 M_\odot$ main sequence star (see Kulkarni, Hut & McMillan (1993) for a discussion of the likely masses of stellar black holes in globular clusters). Finally, the case with a mass ratio of one half is included for completeness.

The upper limits may be extended if one considers smaller velocities at infinity, but one is always restricted by the effect of the local medium - one may calculate the minimum orbital energy a binary may have before it is likely to be disrupted by the gravitational field of nearby stars (see Kochanek (1992) for a discussion).

We use this analysis to put bounds on the mass of the companion of the recently discovered eclipsing pulsar PSR B1718-19 in NGC 6342 (Lyne et al. 1993). This system is particularly interesting because the pulsar has a high magnetic field ($1.5 \times 10^{12}$ G) and a slow spin period (1 s), implying at least two possible evolutionary scenarios (Wijers & Paczyński 1993):

1. An old neutron star captured a low-mass sequence star at most $10^7$ years ago, some mass transfer and hence some spin-up occurred near the time of capture, but true recycling must wait until the companion has evolved to fill its Roche lobe (or some other mechanism leads them to contact). This lends weight to the idea that magnetic fields decay during the recycling phase (Bhattacharya 1992).

2. The neutron star is in fact young, having been formed via the process of accretion-induced collapse of a massive white dwarf.

Wijers & Paczyński (1993) have used the two scenarios to put bounds on the mass of the companion. In the first case, the mass function and likely inclination of the system imposes a lower bound of $0.12 M_\odot$, while assuming the canonical periastron separation at capture of $3 R_*$ imposes an upper bound of $0.35 M_\odot$. In the second case, the mass estimate is put at $0.7 M_\odot$. Thus, future optical observations of this star may enable one to determine the origin of the system.

We may further refine the bounds for the first case as follows. The period of the binary is 6.2 hr and its eccentricity is negligible, so given the mass of the companion, $M_1$, we can calculate its separation, $R_{\mathrm{circ}}$. Assuming that this is twice the separation at capture, $R_{\mathrm{cap}}$, we can compare $R_{\mathrm{cap}}$ with $R_{\mathrm{RL}}$ and $R_{10}$. Table 2 shows that the mass range proposed by

| $m_1$ | $R_{\rm RL}$ | $R_{10}$ | $R_{\rm cap}$ | $R_{\rm circ}$ |
|---|---|---|---|---|
| 0.12 | 5.05 | 6.13 | 8.27 | 16.54 |
| 0.19 | 4.39 | 5.33 | 5.30 | 10.60 |
| 0.25 | 4.06 | 4.91 | 4.08 | 8.16 |
| 0.35 | 3.70 | 4.46 | 2.97 | 5.94 |

Table 2: Predicting the mass of the companion of PSR B1718-19

Wijers & Paczyński (1993) is too wide. We propose that a mass of about $0.2 M_\odot$ seems likely.

If a mass around this value is found, it will lend weight to the present model which proposes a violent period after capture (when some mass transfer may occur) followed by a longer quiescent period with no expansion of the companion except that due to normal evolution, when mass transfer can resume.

## 3 The Model

The model used is outlined in Paper I, and is based on a self-consistent, adiabatic, linearized normal mode analysis of the problem. It is true that for very close encounters, non-linear effects such as mode-mode coupling can become important, but as we will show in a forthcoming publication, (Mardling 1994b) these effects do not alter the general behaviour of the orbit exhibited by the 'linear' system presented here. In fact, the amount of energy extracted from the orbit by the low $l$-modes is not much affected by the inclusion of non-linear terms in the equations representing the system (even when the system disrupts), rather, the main effect seems to be that the higher order $l$-modes not excited by the orbit are excited by those that are, and since these modes have much shorter dissipation timescales than the low $l$-modes, they can dissipate a significant amount of energy in a relatively short time when the low modes are large in amplitude. In particular, for high $l$, the $m = 0$ modes are most excited, and when a system disrupts, these are the modes which become unstable.

We assume a binary is modelled adequately by a point mass and an $n = 1.5$ polytrope, appropriate to tidal capture binaries formed from a compact object and a fully convective low-mass main sequence star.

The scaling used is the standard polytropic scaling (Chandrasekhar 1939) - see Paper I for details. Unless otherwise stated, all models are run with modes up to and including the $l = 4$ $f$-modes.

Given the relatively low velocity dispersions ($\sim$10 km s$^{-1}$) found in globular cluster cores, it is reasonable to take all capture orbits to be parabolic and thus to start all models with eccentricity $e = 1$. The immediate consequence of considering such highly eccentric orbits is that the orbital periods after capture can be prohibitively long, except for the very closest encounters. We have overcome the problem of calculating these orbits (given that the timestep is restricted by the oscillation timescale) by taking advantage of the fact that beyond a certain separation, say $\Delta_c$, the orbit and oscillations essentially become uncoupled. This allows us to compute an analytic solution for both the orbit and oscillation amplitudes for separations larger than $\Delta_c$, and to continue the dynamical calculation for separations smaller than $\Delta_c$ (see Figure 1).

We start by writing down solutions to the decoupled versions of equations (9) and (10)

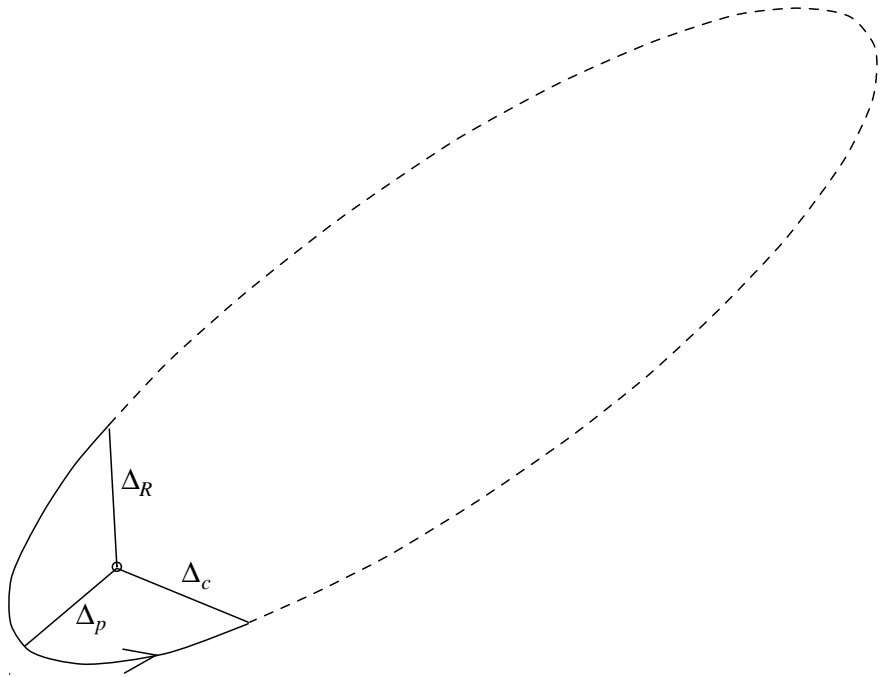

Figure 1: Portion of the orbit calculated dynamically (schematic). The calculation is 'paused' when $\Delta > \Delta_c$ and resumes when $\Delta < \Delta_R$.

of Paper I. These equations are for the orbit and oscillations respectively:

$$\ddot{\boldsymbol{\Delta}} = (1+s)\frac{n}{4\pi}\frac{\partial}{\partial \boldsymbol{\Delta}}\left(\frac{Q}{\Delta}\right) \qquad (4)$$

and

$$\ddot{b}_{\mathbf{k}} + \omega_{kl}^2 b_{\mathbf{k}} = 0, \qquad (5)$$

where $\boldsymbol{\Delta}$ is the radius vector from the centre of mass of the polytrope to the point mass, $\Delta = |\boldsymbol{\Delta}|$, $b_{\mathbf{k}}$ is the oscillation amplitude of mode $\mathbf{k}$ with $\omega_{kl}$ its frequency, $n$ is the polytropic index, $Q$ is the scaled mass, and $s$ is the mass ratio of the compact to the extended star.

The method of solution to equation (4) may be found in Goldstein (1980) and is given by

$$t = T_p + \sqrt{\frac{a^3}{c}}f(\Delta) \qquad (6)$$

and

$$\Delta = \frac{\Delta_p(1+e)}{1 + e\cos(\psi - \psi_p)} \qquad (7)$$

where

$$f(\Delta) = \text{Cos}^{-1}\left[\frac{1}{e}\left(1 - \frac{\Delta}{a}\right)\right] - \sqrt{e^2 - \left(1 - \frac{\Delta}{a}\right)^2} \qquad (8)$$

and $T_p$ is the time at periastron, $\Delta_p$ is the periastron separation, $\psi_p$ is such that $\psi_p - \pi$ is the true anomaly at periastron, $a = \Delta_p/(1-e)$ is the semimajor axis, $c = nQ(1+s)/4\pi$, and $e$ is the orbital eccentricity. This formula is valid for times after periastron, but before apastron. In practice, the orbit is not precisely elliptical until it is effectively uncoupled from the oscillations. Thus in order to calculate $e$ and $a$ (or $\Delta_p$), we take two sets of points $(\Delta, \psi)$ on the orbit near $\Delta_c$, and use the generalized Newton-Raphson method on equation (7) to estimate $e$ and $a$. We can then calculate the time at which the dynamical calculation resumes, $T_R$ (corresponding to $\Delta_R$: see Figure 1). This is given by

$$T_R = T_c + 2\sqrt{\frac{a^3}{c}}\left\{\pi - f(\Delta_c)\right\}, \qquad (9)$$

where $T_c$ is the time at which the dynamical calculation is 'paused'. This time can then be used to determine the oscillatory state of the polytrope at $T_R$, by substituting it in the solution to (5), which is

$$b_{\mathbf{k}}(t) = A_{\mathbf{k}}e^{i\omega_{kl}t} + B_{\mathbf{k}}e^{-i\omega_{kl}t}, \qquad (10)$$

where $A_{\mathbf{k}}$ and $B_{\mathbf{k}}$ are arbitrary constants which may be determined using the values for $b_{\mathbf{k}}$ and $\dot{b}_{\mathbf{k}}$ at time $T_c$. Thus we find

$$b_{\mathbf{k}}(t) = \tfrac{1}{2}\left\{\alpha_{\mathbf{k}}e^{i\omega_{kl}(t-T_c)} + \beta_{\mathbf{k}}e^{-i\omega_{kl}(t-T_c)}\right\}, \qquad (11)$$

where $\alpha_{\mathbf{k}} = b_{\mathbf{k}}(T_c) + \dot{b}_{\mathbf{k}}(T_c)/i\omega_{kl}$ and $\beta_{\mathbf{k}} = b_{\mathbf{k}}(T_c) - \dot{b}_{\mathbf{k}}(T_c)/i\omega_{kl}$.

| $\Delta_p/X_0$ | $e$ | |
|---|---|---|
| 3 | 0.9 | 0.37 |
| 3 | 0.99 | 0.0085 |
| 3 | 0.999 | 0.0002 |

Table 3: Fraction of orbit calculated dynamically

$\Delta_c$ is chosen so that the orbit-oscillation interaction energy (see equation (12) of Paper I) drops below a specified amount. We find that if this is chosen to be about $10^{-6}$, corresponding to $\Delta_c \simeq 42X_0$ (where $X_0$ is the scaled radius of the polytrope), the total energy of the system (see equation (12) of Paper I) after 3000 orbits is conserved to at least one part in $10^4$ and the angular momentum is conserved to at least one part in $10^6$. A solution obtained using this method was compared to that using the full equations, and in particular, the phases of the $b_{\mathbf{k}}$s were compared. The solutions agreed to within one part in $10^6$.

The amount of computing time this method saves can be enormous. Table 3 shows the fraction of the orbit which is calculated numerically for an initial periastron separation of 3 stellar radii and for various eccentricities. Subsequent orbital periods may be even longer, resulting in days of computer time spent calculating one orbit. Another advantage of using this method is that numerical error is reduced, not to mention the accumulated roundoff. Of course, any change affects a chaotic orbit, but we believe that the orbit calculated is shadowed by a true solution to the equations of motion (in the sense of Quinlan and Tremaine 1992). Given that this is true, one is able to examine capture orbits previously inaccessable to study, and these orbits are the very ones most likely to avoid merger and survive to evolve to objects of interest, such as LMXBs.

# 4 Capture Orbits

In this section, we show that the chaotic nature of capture orbits results in a long-time, non-zero average orbital eccentricity, and follows a path in $(\Delta_p, e)$ parameter space which differs significantly from that which corresponds to constant orbital angular momentum, even though the angular momentum transfer to the polytrope remains small. The main consequence of a non-zero average eccentricity is that the orbit can only circularize permanently via dissipation.

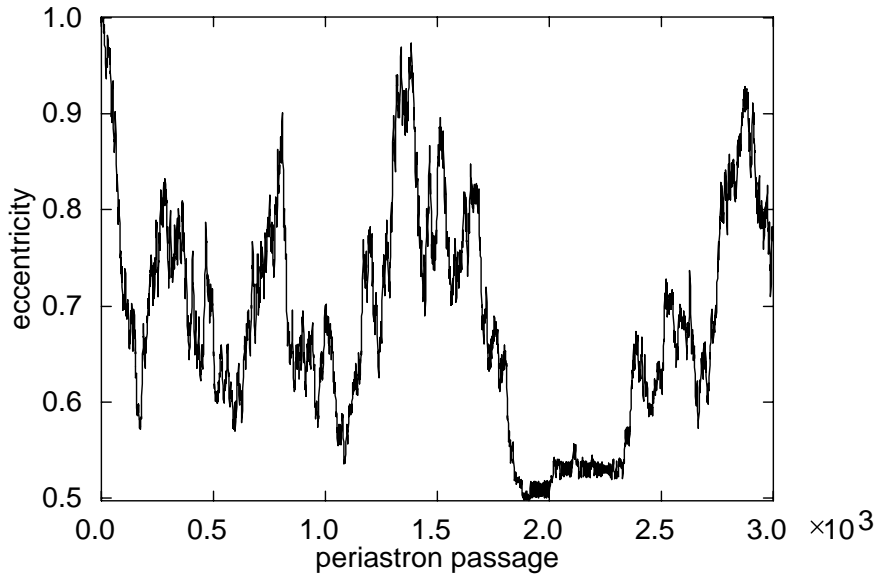

Figure 2: The dynamical evolution of the capture orbit with $\Delta_p^{cap} = 3X_0$. The solution crosses the chaos boundary at around the 2000$^{\text{th}}$ orbit.

We concentrate on orbits with initial periastron separations well away from the lower limit set by the Roche distance. Studies of extremely close encounters such as those of Kochanek (1992), Rasio & Shapiro (1991) and Davies, Benz, & Hills (1991), all of which include non-linear effects (the latter two use Smoothed Particle Hydrodynamics (SPH)), indicate that encounters with initial periastron separations of less than 2.5 stellar radii (for the case with $s = 1$) result in disruption of the stellar envelope(s). This is consistent with results to be presented in Paper III, where we include non-linear effects.

We start by examining the capture orbit with $\Delta_p^{cap} = 3X_0$ and $s = 1$. Wider orbits will be presented, but we use this example to illustrate the main consequences of the model. Figure 2 shows the evolution of this orbit for 3000 periastron passages. The first thing one notices is that the eccentricity does indeed perform a quasi-random walk as predicted by Kochanek (1992), and that there is no permanent circularization. The eccentricity may vary greatly over the course of just a few orbits, or may remain roughly constant for hundreds of orbits, as it does near the 2000$^{\text{th}}$ orbit. This latter feature is extremely important to the evolution of the binary system, and represents a lower bound for the eccentricity, at least in the absence of dissipation. We will examine this in detail shortly. First let us examine the behaviour near capture. Figure 3 shows the first 9 periastron passages of the orbit in Figure 2, where it spends the first 213 years of the

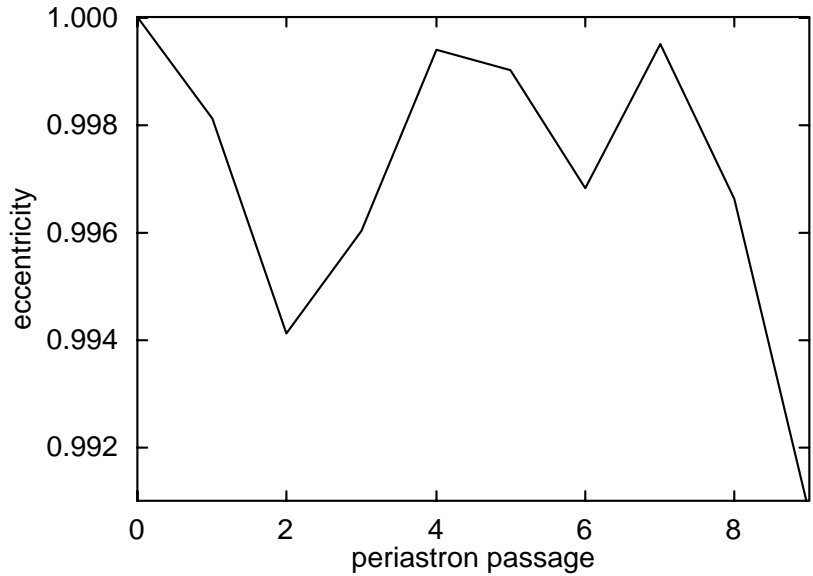

Figure 3: The first 9 orbits in Figure 2.

248 years shown (for $0.7 M_\odot$ stars). Several orbits have eccentricities higher than that of the first orbit following capture, making the binary succeptable to ionization. Given that it survives these first few highly eccentric orbits, the eccentricity quickly reduces and thereafter, attains a maximum eccentricity of 0.975 after periastron passage 1380. Of course, the stochastic nature of the orbit means that it is possible that the eccentricity can come arbitrarily close to unity at some later time, but if dissipation is taken into account, ionization is really only a possibility in the first few orbits after capture.

If only the $l = 2$ mode is included, it is easier for the binary to return most of its energy to the orbit, as Figure 4 shows, again for $\Delta_p^{cap} = 3X_0$. The eccentricity following the $44^{\text{th}}$ periastron passage is 0.9997. If the phase at periastron of the $l = 2$, $m = 2$ mode is such that it returns most of its energy to the orbit, it is unlikely that higher modes, in particular the $l = 3$, $m = 3$ mode would also be in the position to do so. Since the $l = 3$ mode can carry a significant amount of energy (see Paper I), it is unlikely that the binary will be ionized if it survives the first few orbits after capture, when the $l = 3$ mode contains little energy.

Most authors assume that because such a small percentage of the orbital angular momentum is transferred to the stars in the process of circularization, the orbital angular momentum can be regarded as constant and as such, the final separation of the binary will be twice that at capture. This seems a reasonable assumption: for the case we are considering here, by the time the eccentricity has dropped to 0.5, only 5% of the total angular momentum resides in the polytrope. Figure 5 compares the evolution of the present model in the $(\Delta_p, e)$ plane with the curve corresponding to constant orbital angular momentum, given by $\Delta_p(1 + e) = \text{constant} = 2\Delta_p^{cap}$. Some detail is shown in

$$E_{orb} = \tfrac{1}{2} \frac{s}{1+s} Q \dot{\Delta}^2 - \frac{nsQ^2}{4\pi} \frac{1}{\Delta} \qquad (12)$$

and

$$J_o = \frac{sQ}{1+s} \Delta^2 \dot{\psi} \qquad (13)$$

respectively. We can write the orbital energy at periastron (where $\dot{\Delta} = 0$ and the orbital parameters are affected the most) in terms of the orbital angular momentum:

$$E_{orb}(\Delta_p) = \tfrac{1}{2} \frac{1+s}{sQ} \frac{J_o^2}{\Delta_p^2} - \frac{nsQ^2}{4\pi} \frac{1}{\Delta_p} \equiv \alpha \frac{J_o^2}{\Delta_p^2} - \frac{\beta}{\Delta_p}. \qquad (14)$$

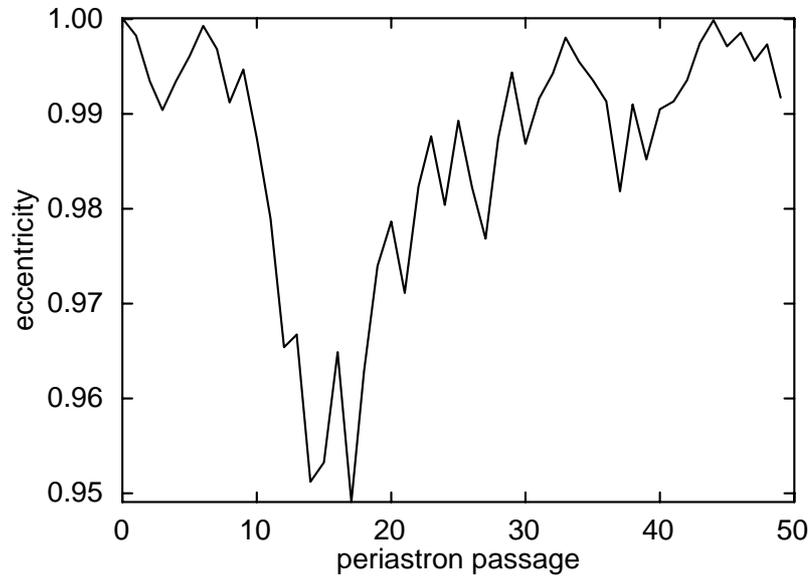

Figure 4: The capture orbit $\Delta_p^{cap} = 3X_0$ with only the $l = 2$ mode included.

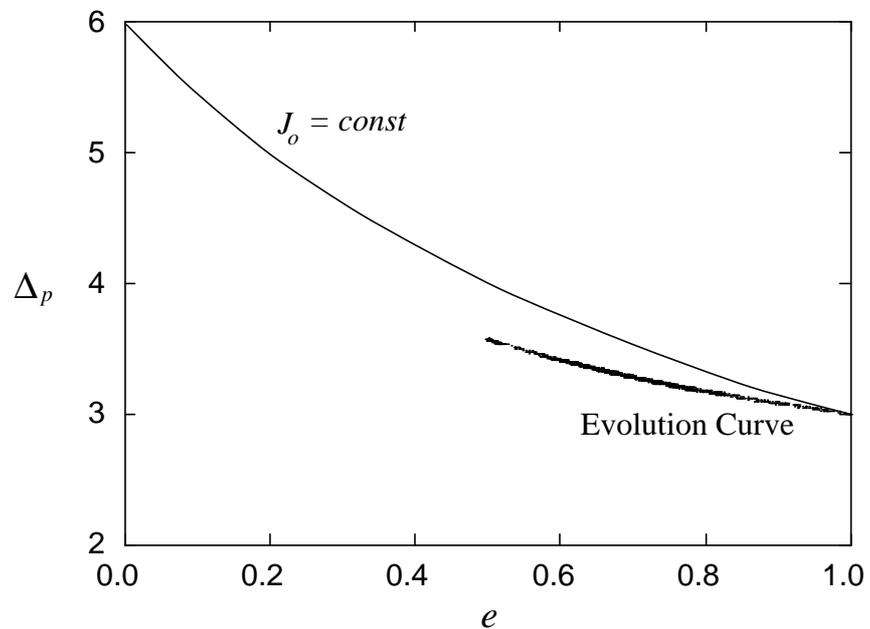

Figure 5: The evolution curve in $(\Delta_p, e)$ space. This is to be compared with the curve along which the orbital angular momentum, $J_o$, is constant: a small transfer of angular momentum to the tides significantly alters the evolution. Here and in the figures which follow, $\Delta_p$ is measured in units of the scaled stellar radius, $X_0$.

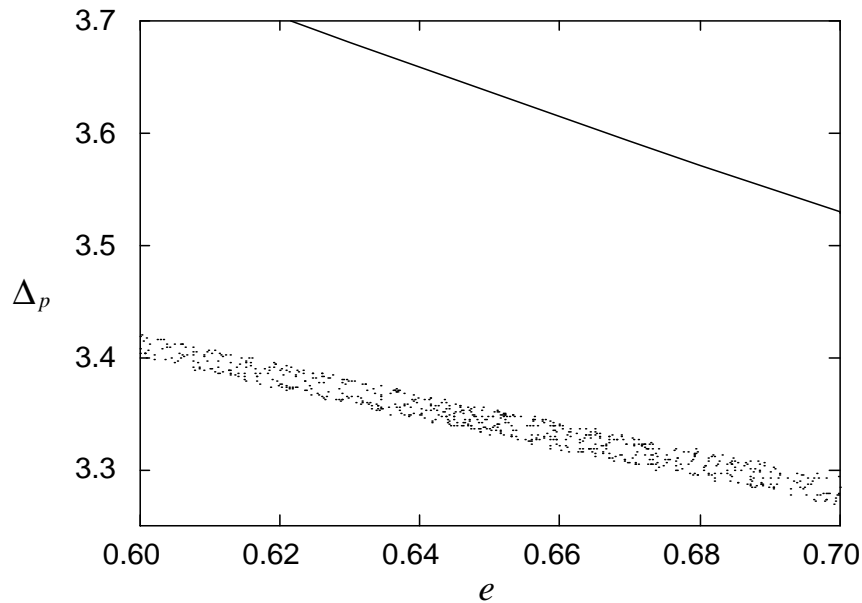

Figure 6: Some detail of Figure 5. Points on the evolution curve are spread between two well defined curves.

Figure 6. Clearly a small angular momentum transfer does not result in a small deviation from this curve. Let us examine why this is the case. The orbital energy, $E_{orb}$, and orbital angular momentum, $J_o$, are (see equations (12) and (13) of Paper I)

Small variations in $J_o$ and $\Delta_p$ result in two contributions to the resulting change in the orbital energy:

$$\delta E_{orb} \simeq 2\alpha \frac{J_o}{\Delta_p^2} \delta J_o + \left( \frac{\beta}{\Delta_p^2} - 2\alpha \frac{J_o^2}{\Delta_p^3} \right) \delta \Delta_p. \qquad (15)$$

When $\delta J_o = 0$, the second terms gives the change in orbital energy appropriate to motion along the constant angular momentum curve in Figure 5. When $\delta J_o \neq 0$, the first term gives the correction. This is negative for angular momentum transfer to the polytrope, and therefore corresponds to the orbit becoming even more tightly bound. Thus, although it is true to say that $\delta J_o/J_o$ is small, the change in orbital energy goes like $J_o \delta J_o$, which is not necessarily small. Equation (15) successfully predicts the change in energy we observe in Figure 5.

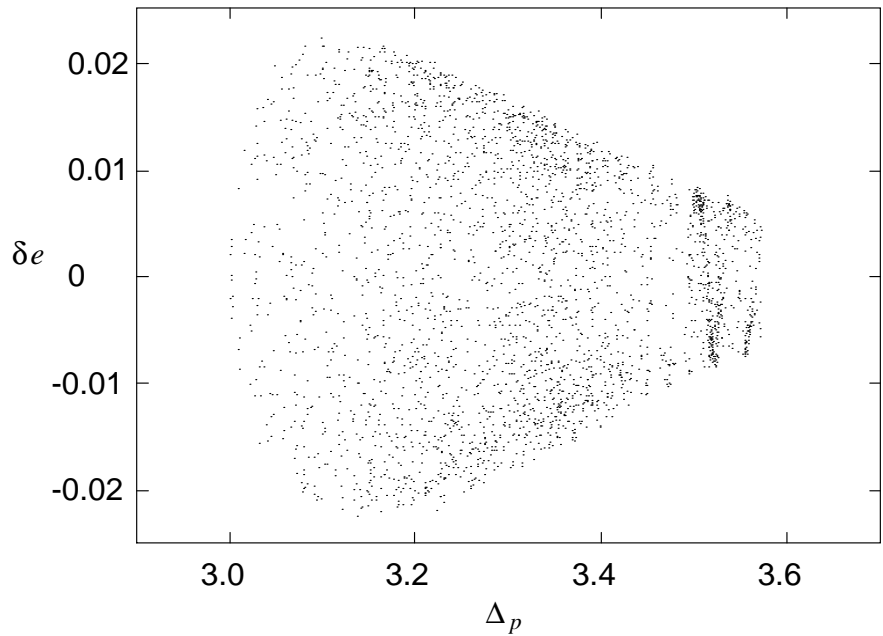

Figure 7: The change in eccentricity (which is proportional to the energy transferred to or from the polytrope) plotted against periastron separation.

It has also been previously assumed (see, for example Kochanek 1992) that as the binary circularizes and the periastron separation increases (through conservation of angular momentum), the amount of energy transferred to the star(s) correspondingly decreases. This is generally the case, although the decrease in not as dramatic as the Press & Teukolsky (1977) model predicts. Figure 7 plots the change in eccentricity against periastron separation. It shows that the energy tranferred to or from the polytrope (which is proportional to the change in eccentricity) may take on any value within a well defined boundary, and that the maximum possible energy transferred actually increases for $3 < \Delta_p < 3.13$, then decreases until at a maximum periastron separation of 3.56 (see later), it is 25% of the value at $\Delta_p = 3.13$. If the Press & Teukolsky model is used, the energy transferred monotonically decreases to 7% of its initial value (which is an order of magnitude smaller than the maximum value here).

We now examine the oscillatory behaviour near the $2000^{\text{th}}$ orbit in Figure 2, which is shown in detail in Figure 8. Here we see behaviour reminiscent of the non-chaotic behaviour examined in Paper I. The solution has crossed the chaos boundary appropriate to the oscillation energy in the polytrope (see Paper I) *and can circularize no further*. Figure 6 shows that for a given value of the eccentricity, the periastron separation is not fixed, but falls between two well defined boundaries. Thus the solution may cross several chaos boundaries depending on how much oscillation energy the polytrope has. Note that the change in eccentricity during this phase can be large, which just reflects the amount of oscillation energy present in the polytrope.

The fact that the solution follows a fairly well defined curve in $(\Delta_p, e)$ space allows us to predict where it will cross a chaos boundary. This will occur when the solution meets a termination point of a chaos boundary corresponding to the amount of oscillation energy present in the polytrope (Paper I). This is shown in Figure 9 which also shows the curve of constant oscillation energy[1] which passes through the intersection of the evolution curve and the 'zero-energy' chaos boundary: neglecting the interaction energy between the orbit and the tides, the oscillation energy at this point is half the maximum oscillation energy the polytrope can acquire. The reason for this is made clear in the next section.

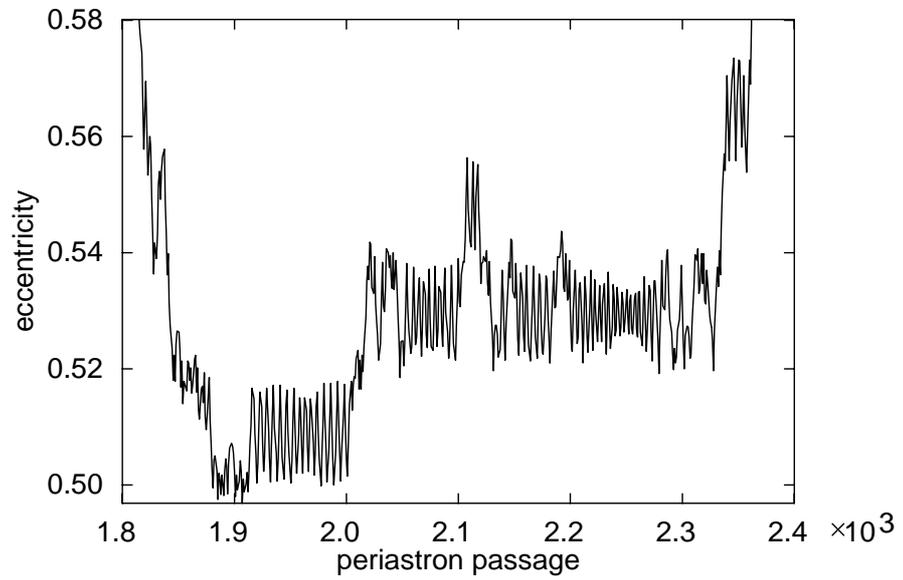

Figure 8: Some detail of Figure 2: the solution has crossed the chaos boundary.

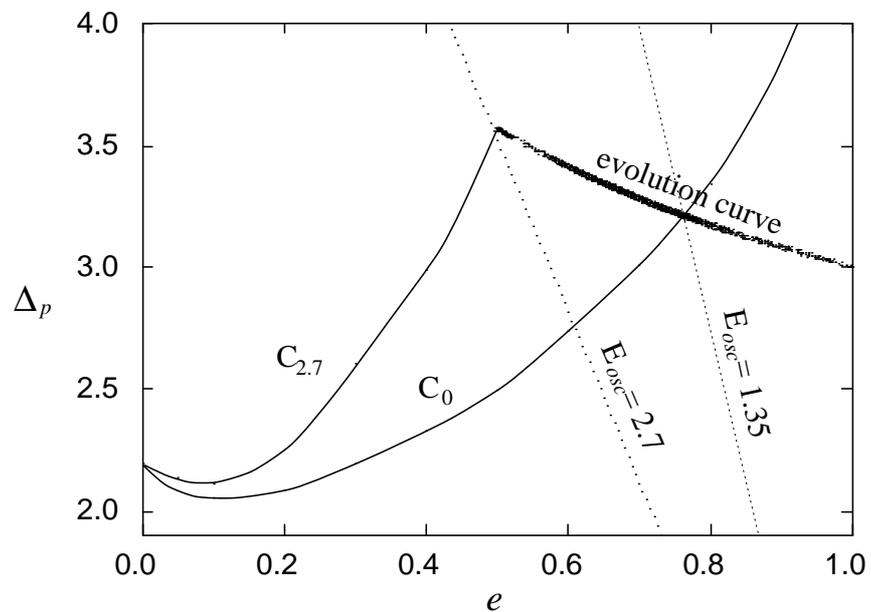

Figure 9: The evolution curve can extend no further than the point(s) where a chaos boundary meets the corresponding curve along which the oscillation energy is constant.

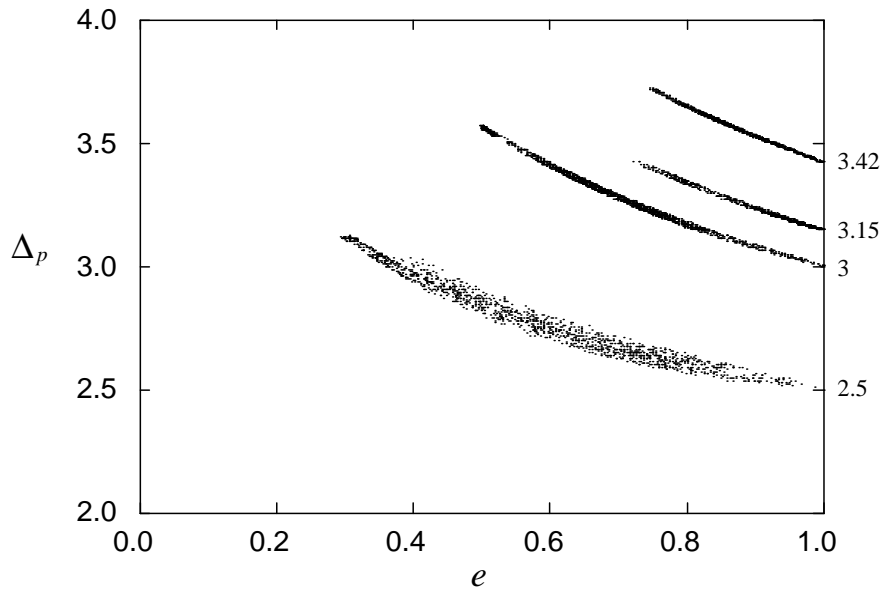

Figure 10: Evolution curves for capture orbits with $\Delta_p^{cap} = 2.5$, 3.0, 3.15, and 3.42. Only the cases $\Delta_p^{cap} = 2.5$ and 3.0 evolved long enough to cross the chaos boundary.

Figure 10 shows the evolution curves for the cases $\Delta_p^{cap} = 2.5X_0$, $\Delta_p^{cap} = 3X_0$, $\Delta_p^{cap} = 3.15X_0$ and $\Delta_p^{cap} = 3.42X_0$. The model with $\Delta_p^{cap} = 2.5X_0$ actually meets the chaos boundary as shown in Figure 11, although this model disrupts when non-linear terms are included. For equal mass ($\sim 0.7M_\odot$) stars, the time taken for these 1500 orbits is about 12 years. The two widest models, which were run for 1000 and 1500 orbits respectively, never circularized sufficiently to meet the chaos boundary.

## 5   Dissipation - The Long-Term Evolution

Various timescales have been estimated for the thermalization of the modes. The longest timescale corresponds to the lowest and most energetic mode, the fundamental $l = 2$ mode, and is thought to be of the order $10^4 - 10^6$ years for a low mass main sequence star (MMT, Ray et al. 1987). If higher modes with much shorter dissipation timescales are able to be excited via non-linear processes, the system may lose energy at a faster rate. However the system loses energy and at whatever rate, we may examine the system's response to this loss.

First of all, consider how the curves of constant oscillation energy will move in the $(\Delta_p, e)$ plane as the system loses energy. These are shown in Figure 12 for zero total energy, together with various chaos boundaries. They will slowly move to the left of the diagram (since for a given oscillation energy, the orbital binding energy will increase), passing through their associated chaos boundaries at points appropriate to the total energy of the system. We illustrate this process schematically in Figure 13. $C_{2.5}$ and $C_{2.7}$ are chaos boundaries for initial oscillation energies of 2.5 and 2.7 respectively. These curves can be marked off according to the total energy a system has if it *starts* at this point in the $(\Delta_p, e)$ plane with the given oscillation energy. If the system loses an amount of energy so that the total energy becomes, say, -0.1, the endpoint of the evolution curve becomes the point at which a (shifted) constant oscillation energy curve crosses the corresponding chaos boundary at -0.1.

---

[1] Such curves are calculated such that any capture orbit reaching a point on them will have the specified oscillation energy. Alternatively, they may be viewed as curves of constant orbital energy (the interaction energy between the orbit and the tides remains small except for extremely close capture orbits).

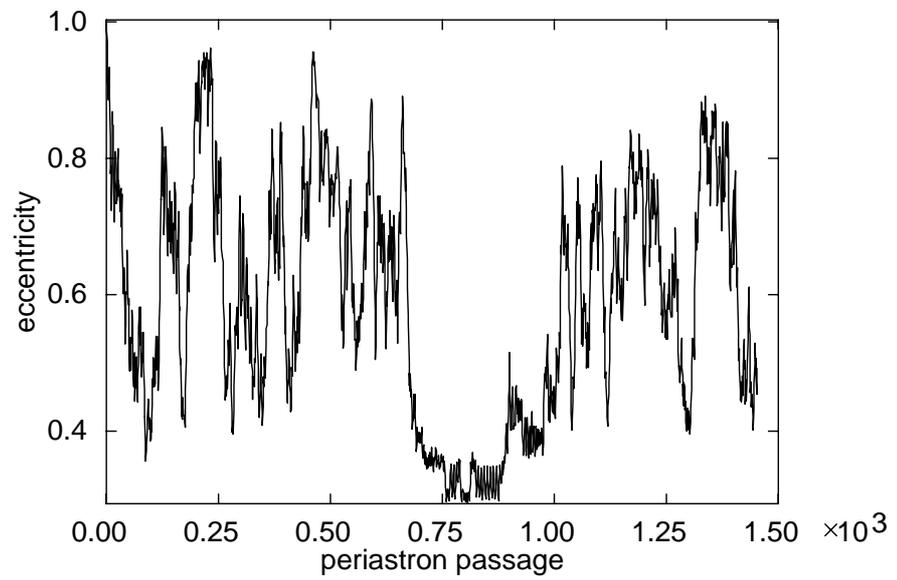

Figure 11: The dynamical evolution of the capture orbit with $\Delta_p^{cap} = 2.5X_0$. The solution crosses the chaos boundary at around the $750^{\text{th}}$ orbit.

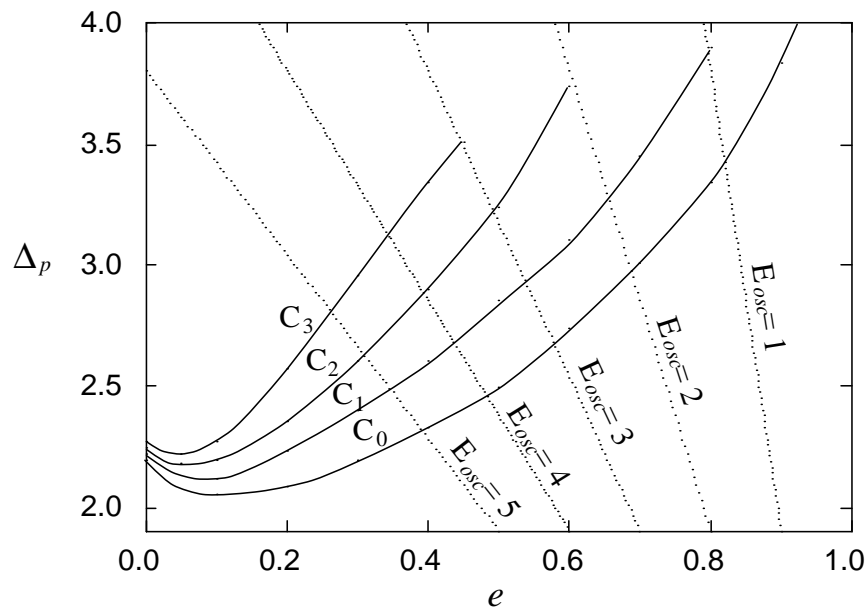

Figure 12: Several chaos boundaries and constant oscillation energy curves. A chaos boundary labeled $C_x$ is calculated by starting the polytrope off with an amount $x$ of oscillation energy. These curves must terminate at a corresponding oscillation energy curve, which are calculated assuming zero total energy. Hence the zero energy chaos boundary terminates at the line along which $e = 1$, which is where all capture orbits are started.

The gist of this argument is that as the system loses energy, the high oscillation energy end point of the evolution curve moves to the right, while the other end of this curve must move to the left by the same amount (the maximum eccentricity the system can achieve reduces). This process continues until the end points meet: this point is necessarily a point on the zero initial oscillation energy chaos boundary. One may see from this argument why it is that the maximum energy the extended star can extract from the orbit is twice the oscillation energy it has at the point where the chaos boundary and the evolution curve intersect.

The evolution curve will also move towards the constant angular momentum curve, since the angular momentum transferred to the stars (which is roughly proportional to the oscillation energy - see Kochanek (1992)) will decrease in time for a given point in the $(\Delta_p, e)$ plane. Thus the end of the chaotic phase will occur at the intersection of the zero energy chaos boundary and the curve of constant angular momentum, and the system need only have dissipated the amount of energy appropriate to this point. This process is illustrated schematically in Figure 14.

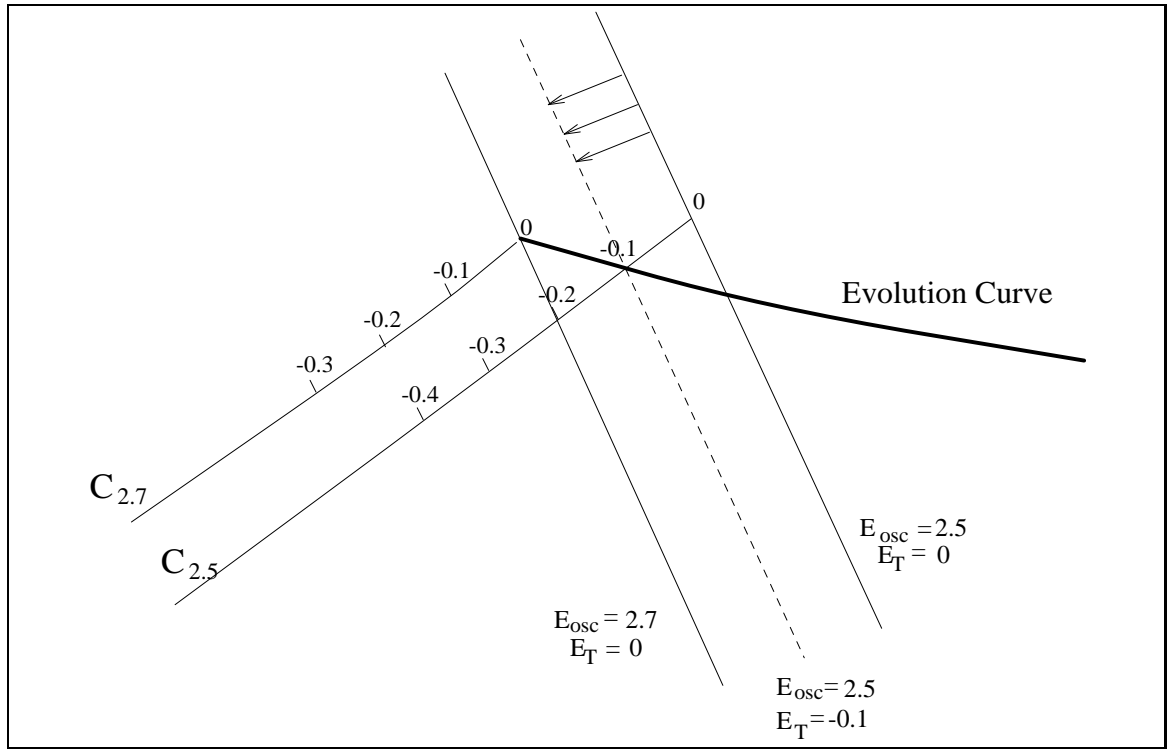

Figure 13: Schematic representation of the effect of dissipation on the evolution. Constant oscillation energy curves move to the left as the system loses energy, so that the evolution curve must now terminate where a (shifted) oscillation curve intersects a corresponding chaos boundary at a point appropriate to the amount of energy lost by the system. Here, $E_T$ is the total energy, and $E_{\rm osc}$ is the oscillation energy.

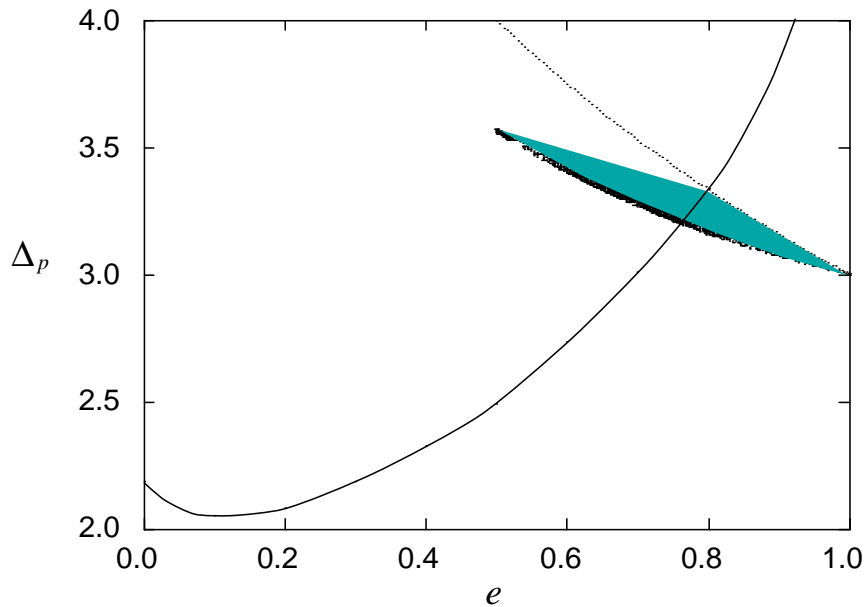

Figure 14: Evolution with dissipation. The evolution curve gradually shrinks towards the 'zero energy' chaos boundary, and moves up towards the curve of constant angular momentum. When the evolution 'curve' reaches the point where the chaos boundary intersects the constant angular momentum curve, the binary has reached the end of the violent chaotic phase. It then begins a long quiescent phase as it moves along the constant angular momentum curve, circularizing *only* via normal dissipative processes.

In effect, the 'zero energy' chaos boundary acts as a kind of attractor, and systems finding themselves at this point in their evolution will no longer behave chaotically, and may circularize *only* via dissipative effects. The tidal energy will from then on always be small, as will changes in eccentricity. As a system loses energy, it will proceed from that point along a curve of constant angular momentum, since the angular momentum transferred to the stars will be truly negligible. The radius of the binary once it has circularized will thus be $2\Delta_p^{cap}$, as is always assumed (neglecting any angular momentum transferred to the stars in the form of a bulk rotation, and in the absence of mechanisms for angular momentum loss such as mass loss, gravitational radiation and magnetic breaking (Hut et al. 1992)).

Hence the zero energy chaos boundary plays a vital role in the evolution of 'wide' capture binaries: it defines the maximum possible tidal energy the star(s) can acquire during the chaotic phase, and it defines the starting point of the long 'quiescent' phase during which the capture binary circularizes only via dissipative processes. We can estimate the length of this quiescent phase as follows. The tidal energy present in a binary started with orbital parameters ($\Delta_p \simeq 3.3 X_0$, $e = 0.8$, see Figure 14) is about $10^{-3}$ (in the present units), while for a circular binary with radius $6X_0$, the tidal energy is about $10^{-4}$ units. The energy the binary needs to dissipate in order to circularize is about 2 units, so taking the lower value of $10^4$ given by MMT for the thermalization time of the $l = 2$ mode (remembering that higher modes contain little energy in non-chaotic orbits - see Paper I), we estimate the circularization time to be of the order of $10^8$ years. This means that tidal capture binaries *are* available as a direct heat source to the cluster core for a substantial part of their life.

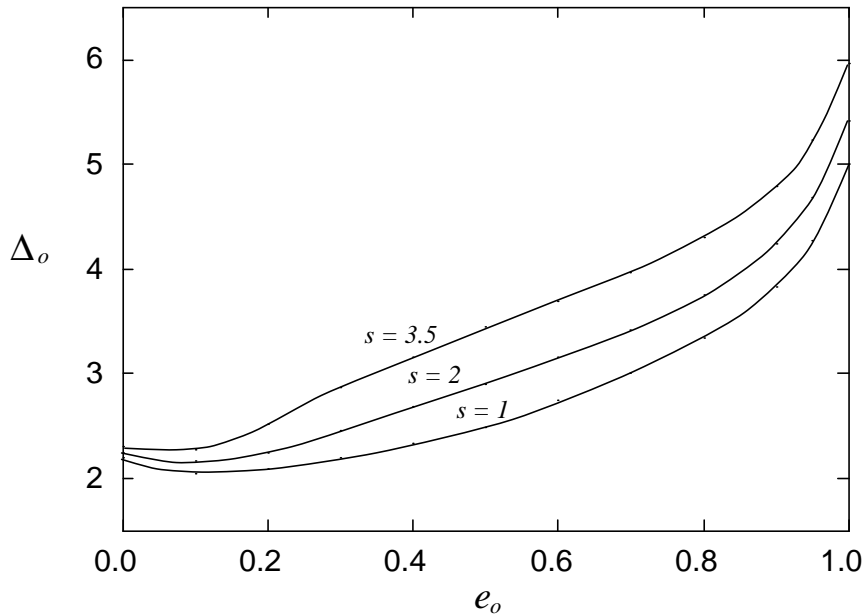

Figure 15: Zero energy chaos boundaries for the mass ratios $s = 1$, 2, and 3.5.

Figure 15 compares the chaos boundaries for cases $s = 1$, $s = 2$, and $s = 3.5$, appropriate to, say, a white dwarf with an equal mass main sequence companion, a neutron star with a $0.7 M_\odot$ companion, and a neutron star with a $0.4 M_\odot$ companion. Referring to Table 1, it is clear that *all* capture orbits are chaotic, at least for stars with a mass distribution something like that of an $n = 1.5$ polytrope. Kochanek (1992) predicted that no stable binaries can form from a neutron star and a main sequence star with a mass below about $0.7 M_\odot$, because the oscillation energy the companion would acquire by the time the binary circularized would be enough to disrupt it. The present analysis shows that if there is a lower limit it will be considerably less than $0.7 M_\odot$, which will be verified if the mass of the companion of PSR B1718-19 turns out to be around $0.2 M_\odot$.

For more centrally condensed stars, the chaos boundary will fall below that for $n = 1.5$ (the outer layers store less energy). Thus it may be possible for a star to be captured but not to behave chaotically, circularizing *purely* via dissipative processes (if it isn't ionized first). In the case of red giants, though, it is likely that $R_{\rm RL} > R_{10}$ (see Section 2 and McMillan, Taam & McDermott 1990), so that stable binaries containing a giant are unlikely to form.

These results have observational consequences. If the above analysis is correct, observations of highly eccentric binaries containing a main sequence star (made apparent by the high apsidal advance associated with such systems) will reveal a very small $\dot{P}/P$, where $P$ is the orbital period, appropriate to the rate at which the system loses energy via normal dissipative processes. On the other hand if the standard model is correct, a much higher value for $\dot{P}/P$ would be measured.

Although the stars can undergo violent interactions in the chaotic phase of their evolution, with probable mass loss for all but the most distant encounters, the question remains whether this phase is long enough to allow the stars to expand substantially. Non-linear effects are likely to hasten this phase. It may be possible for mass transfer from an extended star to a compact object to occur during this violent phase, possibly forming an accretion disc about the compact object. Is it possible that the Rapid Burster (see, for example, Lewin & Joss 1981) is in this phase?

It would be interesting to see whether results from other methods of modelling this problem, such as SPH with a very high number of particles, find that chaotic behaviour is possible.

Finally, of course, the above analysis assumes that the mass distribution of the stars remains unchanged during this phase: any significant change in the mass distribution will affect the chaos boundaries.

# 6 Conclusion

We have deduced that tidal capture binaries do *not* circularize on a short timescale, and that merger is less likely than previously thought. After capture, they pass through a short violent chaotic phase, followed by a longer quiescent phase, the length of which is determined by the rate at which the star can dissipate the modest amount of tidal energy present. The characteristics of the final circularized binary will much depend on the evolution during the chaotic phase, and in particular, whether mass is transferred during this phase.

Tidal capture binaries will be available as a direct heat source for the cluster during their highly eccentric phase, and are less likely to be ejected from the cluster following an encounter with another star.

It remains to be seen, as observational techniques improve, whether a system can be found that will vindicate this model.

Finally, the techniques developed here may be useful for the inclusion of hydrodynamical effects in the N-body simulations of globular clusters.